# Recent advances in inorganic oxides-based resistive random-access memory devices

*Anurag Pritam[a], Ritu Gupta[a], and Prakash Chandra Mondal*,[a]*

[a]Department of Chemistry, Indian Institute of Technology Kanpur, Kanpur 208016, India

Corresponding author: E-mail: pcmondal@iitk.ac.in

**Abstract**

Memory has always been a building block element for information technology. Emerging technologies such as artificial intelligence, big data, the internet of things, etc., require a novel kind of memory technology that can be energy efficient and have an exception data retention period. Among several existing memory technologies, resistive random-access memory (RRAM) is an answer to the above question as it is necessary to possess the combination of speed of RAM and nonvolatility, thus proving to be one of the most promising candidates to replace flash memory in next-generation non-volatile RAM applications. This review discusses the existing challenges and technological advancements made with RRAM, including switching mechanism, device structure, endurance, fatigue resistance, data retention period, and mechanism of resistive switching in inorganic oxides material used as a dielectric layer. Finally, a summary and a perspective on future research are presented.

# 1. Introduction

Our modern-day life demands high-speed data storage, low-voltage operation, continuation access, and low-cost electronic devices. Thus, the ability to store information at a faster timescale is essential for advanced innovative purposes such as cloud computing, data mining, machine learning, the internet of things, and artificial intelligence (AI), to mention a few. For the successful operation of an electronic device, a system needs both temporary as well as permanent memory components, and these requirements are usually accomplished by using semiconductor-based memory devices such as dynamic random-access memory (DRAM), static random access memory (SRAM), and flash memory [1]. SRAMs are considered as one of the high-speed memory devices which are capable of storing information as long as the power is supplied which makes SRM volatile. It utilizes six transistors per bit (6Ts/bit) as shown in **Fig. 1(a)**, and works perfectly for small embedded systems, thus making it unsuitable for the desktop memory system. The DRAM uses the presence or absence of charge to store data and needs continuous refresh cycles to prevent its contents from vanishing. Unlike SRAM, the DRAM also uses only one capacitor and transistor per bit (**Fig. 1 b**) to store charge thus, differentiating between "0" and "1" states. Flash memory or flash storage is a class of memory devices used to transfer and store data between a computer and digital devices. It holds information for a longer period, irrespective of whether either flash device is powered ON or OFF. A flash memory employs the metal oxide semiconductor field-effect transistor (MOSFET) technique to store data by switching the electronic signals shown in **Fig. 1 c**. However, the SRAM is too expensive as its bit cell comprises six transistors, which occupy a large area and consume more power than other technology. Along with the volatility, SRAM is also vulnerable to exposure to radiation [2–6]. The DRAM is slow and requires an enormous amount of power to perform its function. DRAM has a strong chance of data leakage, which means the cell is refreshed constantly and displays an added complexity in the memory device. Flash memory is limited as it requires a high electric field to accomplish erase and rewrite operations [7–9]. Therefore, researchers have been effectively probing the charge-independent memory devices to revolutionize the existing memory hierarchy. The classification of memory devices is shown in **Fig. 2**. These evolving information storage technologies target to combine the non-volatility behavior of flash drives along with the storage density of DRAM and the **Erase-Rewriting** speed of SRAM, which will make them a fascinating replacement for future memory technology. The ideal memory device should possess necessary features such as excellent data retention time, superior scalability, low operating power, and less energy consumption. However, it is challenging to find any memory devices with all these ultimate features.

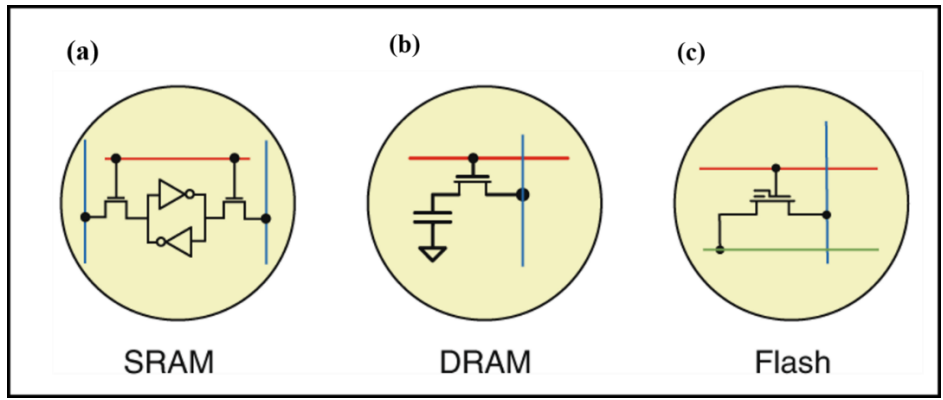

*Fig.1 Schematic configuration of (a) SRAM (b) DRAM (c) Flash memory*

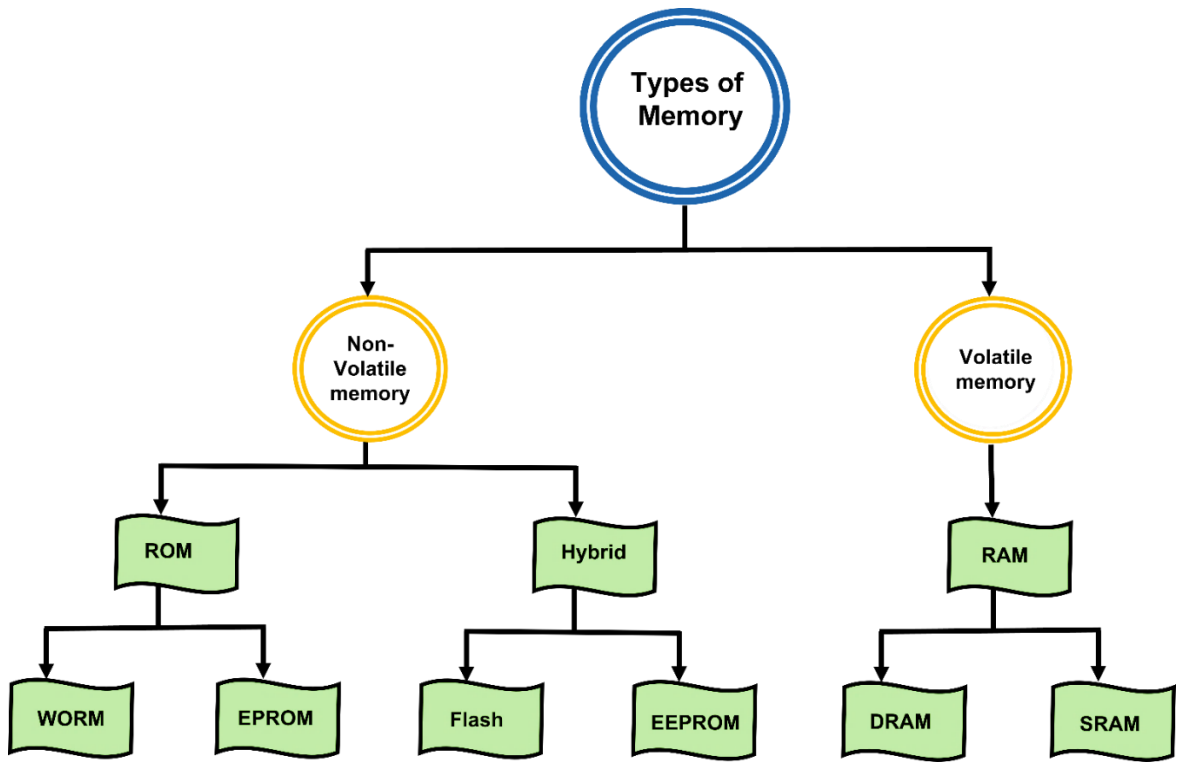

*Fig.2 Classification of memory technology*

Recently researchers have come up with novel resistance-based memory technology which can meet a part of these ideal qualities. In place of a charge storage mechanism, these class of memory devices store information in response to changes in the arrangement of atomic positions of the ferromagnetic layer, this change in the orientation leads to a change of resistance of the system and is termed resistive memory devices, occasionally the resistive memory devices are also known as memristive devices, due to their association with the idea of the memristive system. A few notable examples of this class of memory devices are spin-transfer torque magnetoresistive random access memory (STT-MRAM), Phase change memory (PCM), and Ferroelectric random-access memory (FeRAM), and resistive random-access memory (RRAM) [10,11]. A brief idea of the same is provided in the next paragraph.

Spin-transfer torque magnetoresistive random access memory (ST-MRAM) is a category of nonvolatile solid-state information storage device, where the magnetic tunnel junction (MTJ) plays a decisive role in the storage of information/data. The MTJ comprises two ferromagnetic layers, one has a free magnetic alignment, and the other has a fixed alignment. The two ferromagnetic layers have been separated by an insulating dielectric tunnel barrier of the thickness of 4 to 6 nm shown in **Fig. 3a**. If the direction of magnetization is identical in the reference layer as well as in the free layer, then the MTJ cell is known to be in a low resistance state (LRS), and when the direction of magnetization is antiparallel, then it is in high resistance state (HRS), respectively. The data recording in STTMRAM is executed by changing the magnetic alignment of the insulating layer by using spin-polarized current, and subsequent difference in the resistance of the MTJ cell is used for information storage. STT-MRAM is a more appropriate technology for upcoming high-end memory devices due to its unique features such as simple architecture, low energy consumption, high scalability, and high-speed operation. ST-MRAM displays much better endurance compare to that flash drives with excellent data preservation properties. Irrespective of all these outstanding properties, the ST-TMRAM also faces some severe issues in small tunneling magnetoresistance ratio, lack of suitable material for fabrication, critical size, thermal instability, etc [12–17]. Phase change memory (PCM), also known as chalcogenide RAM (CRAM), is a nonvolatile memory device that stores information at a nanometer scale. The essential feature of any memory device is that it should allow the storage and fetching of the data. PCM memory device comprised of a phase change material sandwiched between two conducting electrodes. PCM device writes data by recording the change in electrical resistance of phase change material when it switches from the ordered crystalline phase (low resistance state "1") to the strongly disordered amorphous phase (high resistance state "0") and vice versa, shown in **Fig. 3b**. The data retention duration of PCM is very high as it can store information for 8-10 years at a room temperature of 25 ºC, which could make PCM be effectively used as Flash memory whose operating speed is almost fast as compared to volatile memory devices like DRAM [6–9,18]. Irrespective of many valuable properties and applications, PCM also possesses a few challenges which need to be addressed. The first one is the need for high programming current density ($10^7$ A/cm$^2$) to match the level of a diode. Unarguably the major concern associated with PCM is its threshold voltage drift and long-term resistance [11,18–25]. By obeying the power law, resistance of the amorphous phase increases slowly with the applied voltage, thus restricting it from the multilevel operation and hindering the regular two-state operation if the threshold voltage is raised above the designed operating value.

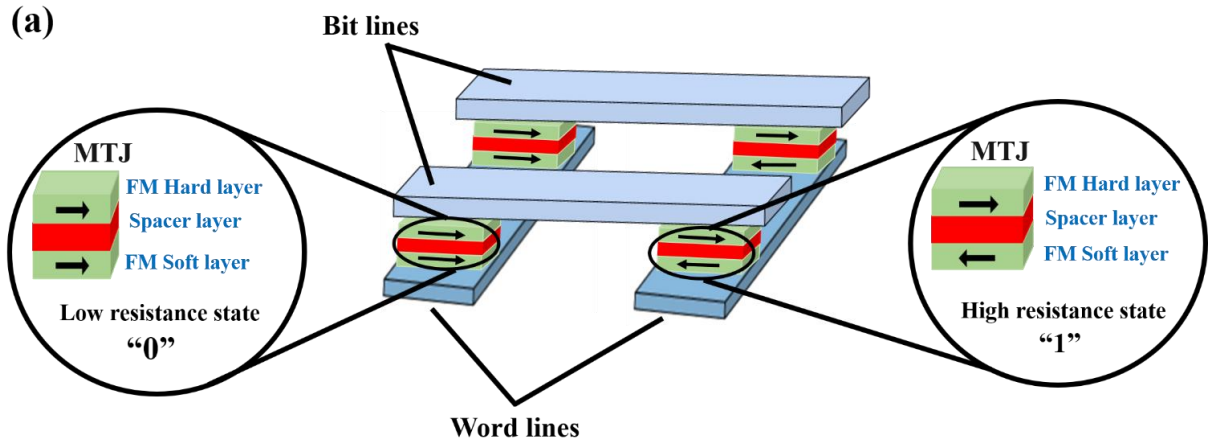

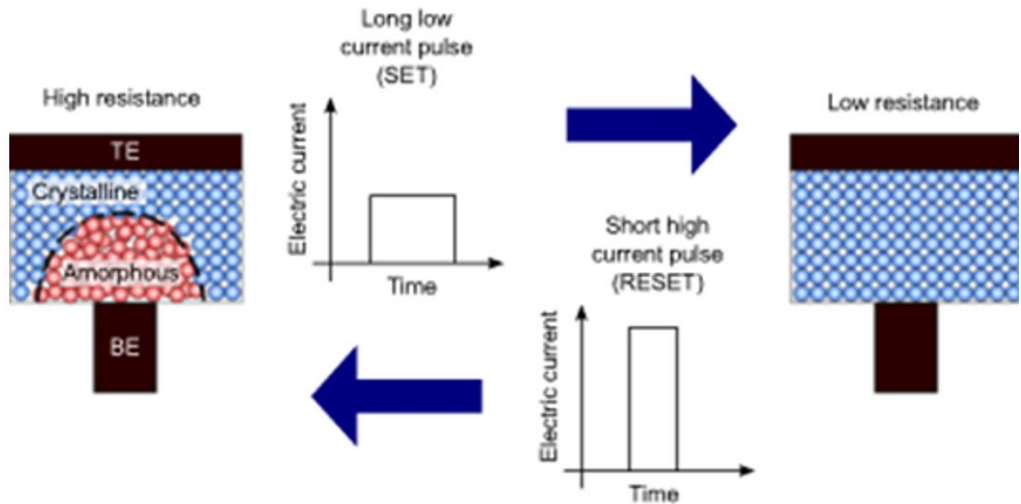

*Figure 3. Schematic representation of MRM and PCM.*

Apart from PCM and MRAM, researchers have also explored another class of nonvolatile memory devices known as resistive random-access memory (RRAM) devices. RRAM possesses many significant advantages over other memory storage devices such as RRAM can be easily scalable, easy fabrication technique, high-speed operation, very high erase/write cycle, low power utilization, ultra-fast speed, and long data retention duration. This forces the researchers to carry out an extensive amount of research on eliminating the flaws and enhancing the properties of RRAMs so that they can offer an economical solution to future memory devices. The comparisons of properties and configuration of different types of RAMs are depicted in **Table.1**.

**Table.1** *Comparison of different memory device's performance parameters*

| Operational Parameters | SRAM | DRAM | NAND FLASH | NOR FLASH | PCM | STT-MRAM | RRAM |
|---|---|---|---|---|---|---|---|
| Cell area | $> 100F^2$ | $6F^2$ | $< 4F^2$(3D) | $10F^2$ | $4$–$20F^2$ | $6$–$20F^2$ | $< 4F^2$(3D) |

| Cell element | 6T | 1T1C | 1T | 1T | 1T(D)1R | 1(2)T1R | 1T(D)1R |
|---|---|---|---|---|---|---|---|
| Voltage | <1V | <1V | <10 V | <10 V | <3 V | <2 V | < 3 V |
| Read time | ~1 ns | ~10 ns | ~ 10 μs | ~50 ns | < 10 ns | < 10 ns | < 10 ns |
| Write time | ~1 ns | ~10 ns | 100μs–1ms | 10μs–1 ms | ~50 ns | <5 ns | < 10 ns |
| Write energy | ~fJ | ~fJ | ~10 fJ | ~10 fJ | 100 pJ | ~0.1 pJ | ~0.1 pJ |
| Retention | N/A | ~64 ms | >10 y | >10 y | >10 y | >10 y | >10 y |
| Endurance | $> 10^{16}$ | $> 10^{16}$ | $> 10^4$ | $> 10^5$ | $> 10^9$ | $> 10^{15}$ | $\sim 10^6$–$10^{12}$ |
| Multibit capacity | X | X | √ | √ | √ | √ | √ |
| Non-volatility | X | X | √ | √ | √ | √ | √ |
| Scalability | √ | √ | √ | √ | √ | √ | √ |

To fabricate high-performance RRAMs, numerous kinds of materials have been extensively explored, starting from quantum dots, hybrid composite, layered material, 2D material, ferromagnetic material, and organic and inorganic materials. In this review work, we will be focusing on inorganic metal oxides compositions for RRAMs application, as these materials possess lots of advantages like fast switching cycle, less fabrication cost, low power consumption, long term stability, and many more [26]. Here we briefly discuss their operational mechanisms.

## 2. Fundamentals of RRAM devices

Over the years, resistive random access memory (RRAM) has been extensively studied owing to its unique properties and easy fabrication process, which proves to be the perfect replacement for transistor-based flash memory devices[5,6,27–29]. The RRAM is a kind of memory device that operates on change in resistance of the material under the application of electric potential. This phenomenon has been widely studied in numerous oxide materials due to their adaptability to the complementary metal-oxide-semiconductor (CMOS) nanofabrications [4]. RRAM possesses a considerable advantage over other exiting technology in terms of high fatigue resistance, high polarization switching ($10^{12}$ cycles), low operating voltage (~ ±1 V), ultrafast processing speed (< 1 ns), and can be further scaled down to compact size, thus can be utilized as the next generation Nonvolatile memory technologies [25,30–33]. The structural configuration of resistive RAM is based on capacitor type metal insulator metal (MIM) structure shown in fig.4(a). The insulator layer used in the RRAM device is usually made up of oxide elements like Tantalum (Ta), Titanium (Ti), etc.; some of the 2D materials and chalcogenide have also been reported to be used for the same [6,27,32,34–40]. Stacking of the RRAM devices can be available in the form of a single layer and as well in the form of multilayered stacking depicted in **Fig. 4b**.

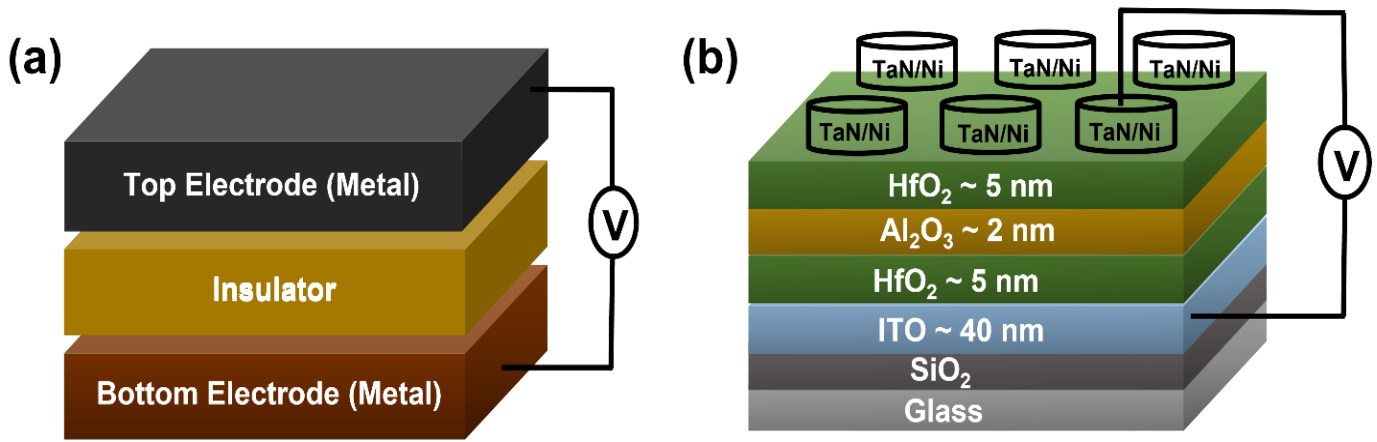

Fig.4 Schematic representation of MIM structure embedded in RRAM (b) multilevel switching in RRAM.

## 2.1 Operating principle of RRAMs

RRAMs usually operate by exploiting the reversible alteration in the electrical resistance of the insulating oxide layer. Usually, the conductive path plays a vital aspect in the operation of resistive RAM. The conducting path usually consists of conducting filaments, which form in the insulator usually due to defects, strain, and mobile oxygen vacancies, and its process of formation is discussed here.

Initially, the RRAM used to be in a high resistance state (HRS), and an externally applied voltage transforms it into a low resistance state (LRS) usually happens due to the formation of conductive filaments in the switching layer. These conducting filaments reduce the device's net resistance, thus providing an easy pathway for the flow of current across the device which eventually led to the soft breakdown of metal/insulator/metal (MIM) structure, usually known as a process of electroforming and the voltage at which electroforming occur is known as forming voltage ($V_f$). Forming voltage strongly depends upon the thickness of the oxide layer and the area of the cell. The mechanism of switching resistance from LRS to HRS is known as SET, and the device is known to be in an ON state, however when the device is switched back to LRS from HRS, known as RESET operation, the resistance of the device drastically increases due to breaking of the conductive filaments, which results in blocking of supply of current, and eventually, the device turned to be in OFF state. Depending on this SET and RESET state, the memory device reads a bit "1" or bit "0" thus, responsible for the resistive switching phenomena in RRAM shown in **Fig. 5** [27,32,34,36,41–48].

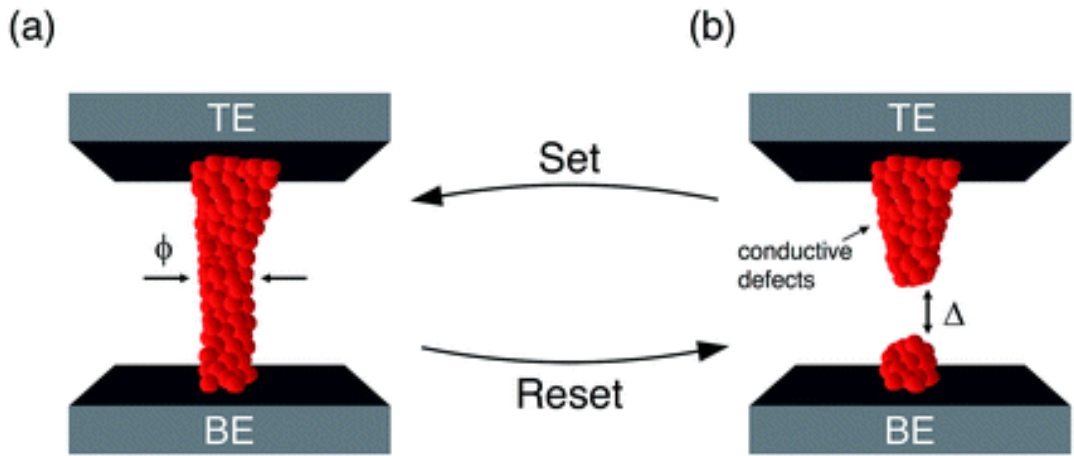

*Fig.5 Depiction of SET and RESET states of the RRAM device*

Sometimes many **S**ET/**R**ESET (**S/R**) operations can damage the electrical component of the device, which usually comes into the picture due to the construction of uncontrolled conductive filaments; this can be checked by the current limit known as compliance current which protects the device from the further damage. For RRAM to function appropriately, a forming process is needed to be done during electrical measurements. Generally, in the virgin device, the conduction of current is hindered due to the presence of an adequate number of defects in the insulator layer, which can be corrected by the soft breakdown of the insulator layer by applying a high forming voltage. This applied voltage increases the concentration of defects in the dielectric layer, which eventually helps to trigger the high current during routine operation. However, some RRAM devices are forming free as they don't have to go through the high voltage forming process because the defects can be introduced in the insulator layer of these devices at the time of fabrication only, thus allowing an easy pathway for the conduction of current during regular SET/RESET operation.

On basis of the polarity of externally applied voltage, the resistive RAM is further categorized into three different types of switching modes: unipolar, bipolar, and single resistive type switching mode. In unipolar switching, the SET and RESET phenomena of the data storing device are sovereign of the polarity of applied voltage bias, and switching of the resistance states occurs by providing a voltage of the identical polarity but with a contrasting amplitude. The current-voltage relation of the same is shown in **Fig 6**. In unipolar switching, joule heating plays a crucial role in breaking the conducting filaments during the RESET process. Moreover, in bipolar switching, the SET and RESET mechanism usually depends on the polarity of the applied voltage; switching from HRS to LSR is taken care of by the same polarity, whereas the opposite polarity shifts the RRAM cell back to its HRS. Rupturing of conducting film in bipolar switching is performed by migration of defects, oxygen vacancies, and also well supported by joule's heating. In single resistive switching, the memory device is unable to trace back to HRS once the resistance changes to LRS, this kind of device is also known as write once read many times (WORM) memory [1,3,6,19,37,49–59].

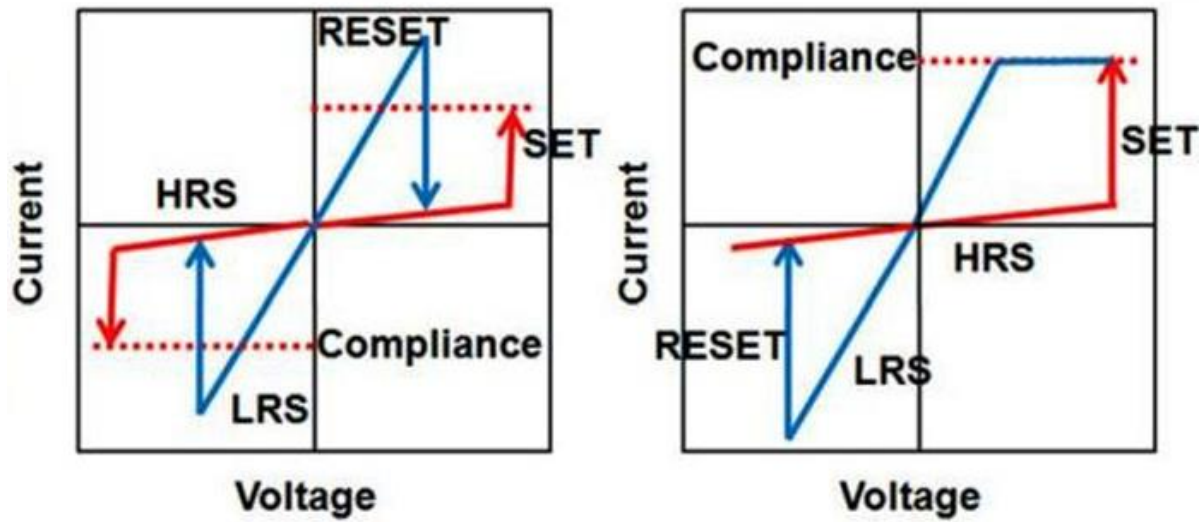

**Fig. 6 Unipolar and Bipolar Switching in RRAM device.**

## 2.2 Types of switching mechanisms in RRAMs

Based on the fabrication process, numerous switching mechanism can be exploited in RRAMs few of them is discussed below which are (a) electrochemical switching, (b) thermochemical switching, and (c) valence charge switching.

*<u>Electrochemical switching</u>*: The switching within this group relies on the movement of charge under the effect of electric field and is termed as conductive bridge RRAMs or electrochemical metallization RRAMs (ECM-RRAM). The switching mechanism in these ECM-RRAMs is activated by the shift of positively charged ions under the external electric field and by creating conducting filaments in the dielectric layer of MIM. In ECM-RRAM, the MIM consists of a metallic electrode such as Cu or Ag, an inert electrode usually made up of platinum, and an electrolyte that is sandwiched between them [1,3,4,11,58]. The formation of conductive filaments determines the SET/ON state due to the drifting of highly mobile $Ag^+$ on the insulating layer finally reducing at the counter electrode. Furthermore, the RESET process is performed by rupturing of conductive filament, which eventually hinders the current flow in the MIM, depicted in **Fig.7 a**.

$$Ag_2 + (Ag_2S) + e^- \text{-----------------------} Ag$$

**Thermochemical switching:** In this switching mechanism, the redox reactions are thermally induced, thus leading to the variation of local conductivity and resistivity of the material and known as thermochemical RRAMs (TCM RRAMs). Due to the thermal behavior of the fundamental process, the switching mechanism between different resistive states is unipolar. The SET and RESET operation is carried out by heat-driven formation rupture of the conductive filament in the non-conducting layer of the MIM structure [1,35].

**Valence change switching**: This type of switching mechanism is dominant in the devices based on mobile donor type defects (like oxygen ions). In presence of applied voltage, the oxygen atom is released from the insulating layer and migrates towards the electrode leaving behind oxygen defects, depicted in **Fig. 7c** which eventually lead to the formation of defect-induced conducting filaments. Usually, in the VCM device, the transition metal oxide (TMO) with inert metal like Pt is utilized as an electrode to obstruct the drift of ions from the electrode. A VCM system can be can further distinguished into two groups filamentary switching followed by interface switching. In filamentary VCM devices, the RS behavior occurs due to the creation of vacancy-based filament, whereas in interface type VCM, the switching is carried out through the thickness modulated tunnel barrier mechanism. The interface type VCM displays a stable switching mechanism compared to filamentary type VCM devices [35,58,60].

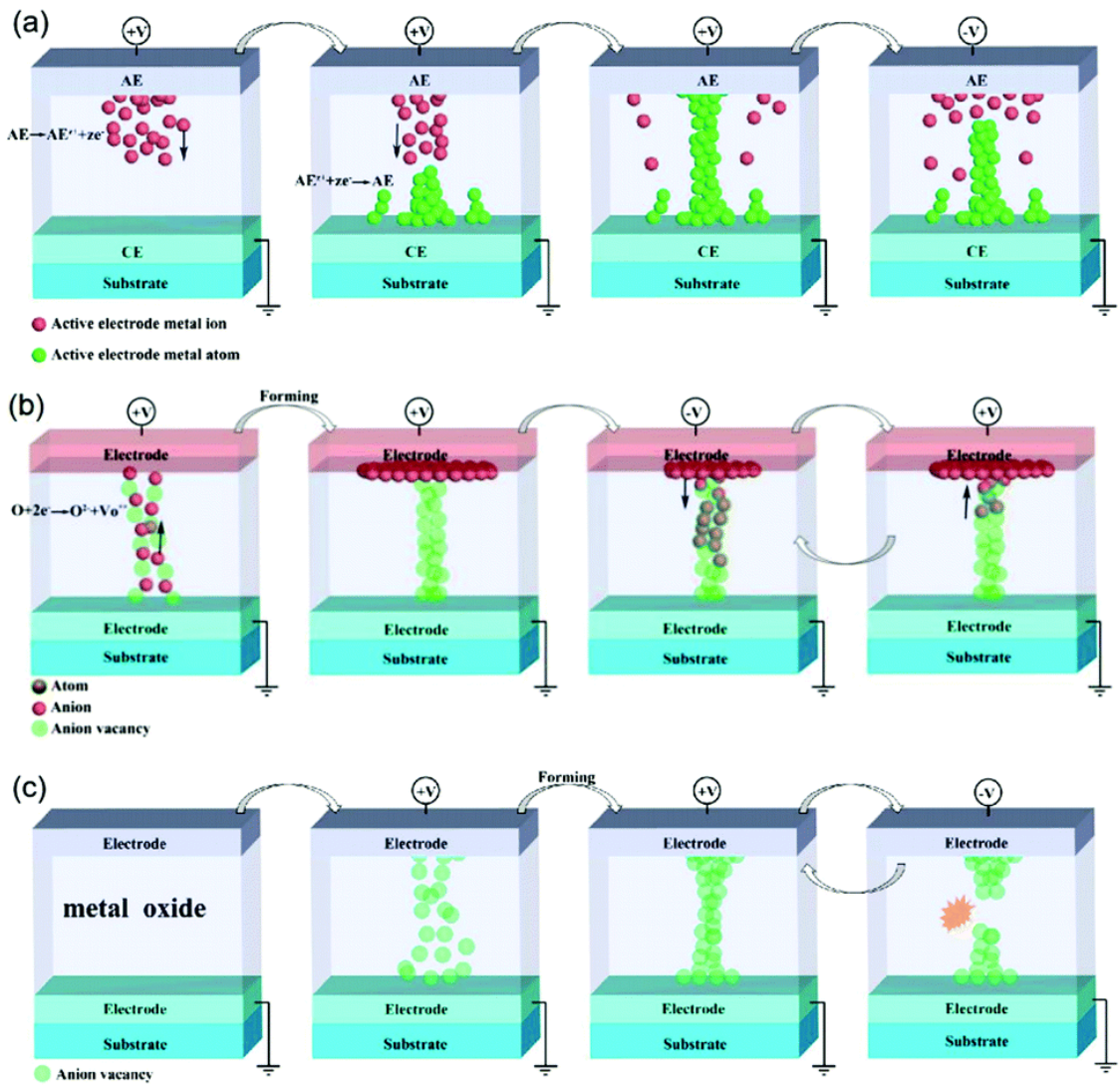

*Figure 7 Schematic description of electrochemical switching (10.1039/D0TC03668D)*

## 2.4 Performance index of RRAMs

The switching mechanism of the RRAMs strongly depends on some figure of merits such as endurance, speed, data preservation, power consumption, switching time, operating voltage, variability, etc. This section will discuss these factors in detail and highlight some best work reported on RRAMs.

### 2.4.1 Endurance:

Operating of RRAMs involves multiple switching from HRS to LSR and vice versa; each switching process introduces some kind of permanent damage to the RRAMs, which eventually degrades their performance, this degradation is known as endurance. The endurance of any memory device is known by executing a series of current-voltage sweeps in a resistive switching cell and extracting the value of resistance for both LSR and HRS by applying a small amount of read voltage, normally at 0.1 V. Generally, the endurance of around $10^{12}$ has been reported in RRAMs device comprise of Ta/TaO$_x$/TiO$_2$/Ti stacks and Pt/Ta$_2$O$_{5-x}$/TaO$_{2-x}$/Pt. Many groups have observed that if the resistive layer of the memory device is made up of oxygen-rich Tantalum, it will display high endurance value [1,58,61]. Several aspects influence the total endurance of the memory device. The first one in this list is temperature; it has been observed that the endurance value of the RRAMs is drastically decreased with the increasing temperature, second factor is the choice of the writing mode, as the endurance value is very high in writing a single byte compared to page write. In page write, the charge pump is distributed over multiple bytes, whereas in programming a single bit, the pumping of charge is very much focused on a single byte, thus leading to the cell's wear and tear. The third and final point is the designing issue of the device, the researcher should explore numerous ways to design a memory device of excellent endurance for better performance.

### 2.4.2 Data retention:

Data retention of RRAMs is defined as the period over which the HRS and LRS of the memory device remain stable after undergoing SET and RESET switching, irrespective the device is electrically powered ON or OFF. The state retention of RRAMs is known by measuring the Current vs. time curve by applying constant voltage stress for both HRS and LRS. The data retention time in LRS is usually low, and the reason behind this is the set voltage induced atomic rearrangement. In contrast, in HRS, the retention time is generally in the natural state of the MIM structure, and RRAMs will persist in the same state if no external voltage bias is applied. Data retention bank on temperature, usually decreases with the increasing temperature, possibly due to the repeated change in the atomic positions. The average data retention period of RRAMs-based on inorganic heterostructures is approximately 10 to 15 years.

### 2.4.3 Homogeneity:

For the successful operation of the RRAMs device in day-to-day life, it should execute a stable resistive switching. One of the primary reasons that prevent it from the further application is the poor homogeneity issue. Few factors responsible for the same are switching voltage, fabrication process, and the resistance of RRAMs in its LRS and HRS, thus creating a high degree of variability. Among all, the material used during the switching process plays a crucial role in variability in RRAMs devices. As we know, the formation of conductive filaments usually creates the LSR, and the conduction of current in LRS generally displays ohmic behavior. Still, in some cases, the non-ohmic or tunneling type behavior is also observed. Whereas HRS is formed by rupturing of the conductive filament of MIM structure, and the conduction mechanism is usually governed by Schottky type, pool Frenkel type, and space-charge limited current (SCLC), makes it challenging to forecast the mechanism of conduction in HRS, eventually make it relatively challenging to portray the variability of HRS compare to LRS. Device variability along with the conduction mechanism in HRS depends on several factors including processing condition, dielectric nature and strength of the insulating layer, defects mobility, and interfacial properties of the electrode. There are a couple of well-established materials, such as $Ta_2O_5$, $Al_2O_3$, $TiO_2$, $HfO_2$, etc., which are successfully used for resistive switching in RRAMs devices [58,61].

## *2.4.4 Speed:*

A switching time between two states plays an important factor that defines the efficiency of the RRAMs device. The extensively used NAND Flash can display a switching time of 100 µs. In contrast, the newly emerged technology such as PCRAM and ST-TRAM has been reported to achieve operation speeds of 2.4 ns and 10 ns, respectively. Among different types of RRAM, the conductive filament type RRAMs is reported to exhibit the switching speed of 300 ps, thus making them one of the fastest and most efficient memory devices. Several reports suggest that the switching in the sub-nanosecond region is infrequent and very difficult to achieve. The switching speed of more than 120 ps was claimed in $Ta_2O_5$ based RRAM devices in 2011 is by Torrezan et al. [49,62]. A couple of other researchers have also observed the SET and RESET switching time less than 500 ps in Pt/SiO2-based memory devices, an impressive switching speed under 85 ps has been achieved in nitride-based memory devices [60,62–65]. All this research concludes that the switching rate strongly banks on the current-voltage (I-V) measurement setup of the memory devices.

## *2.4.5 Power utilization:*

The vast scale application of memory devices depends on the power consumption during switching between SET and RESET. Among all other class memory devices, the power consumption of RRAM is usually very low. Power utilization of 0.1 pj/bit has been reported for TiN/Hf/HfOx/TiN structured RRAM device, and it varies between 0.1 to 7 pj/bit for Al/Ti/$Al_2O_3$/s CNT structured memory device. Hung et al. show that the RRAM memory device consists of ITO/TiOx/Ag nanoparticle/ TiOx/ AlTiOx/FTO. The switching between the two resistive states in the reported device is usually governed by the construction and destruction of conductive filaments in the area between silver nanoparticles and the electrode (FTO and ITO). Power consumption in the SET and RESET process is around 10 10 µW and 0.65 µW, thus displaying low power consumption during operation [33]. The I-V plot and resistive switching mechanism of the device during SET and RESET are shown in **Fig 8 and Fig 9**.

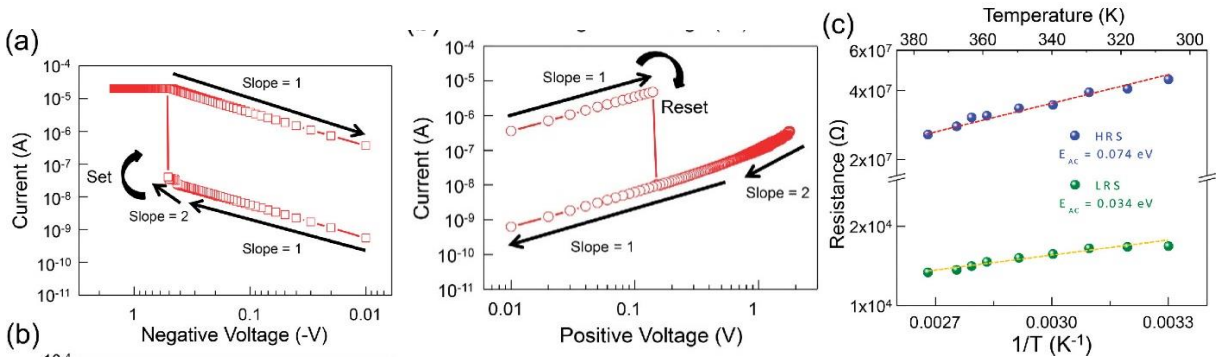

**Figure 8**. I-V characteristics of a memory device (need to take permission form F8)

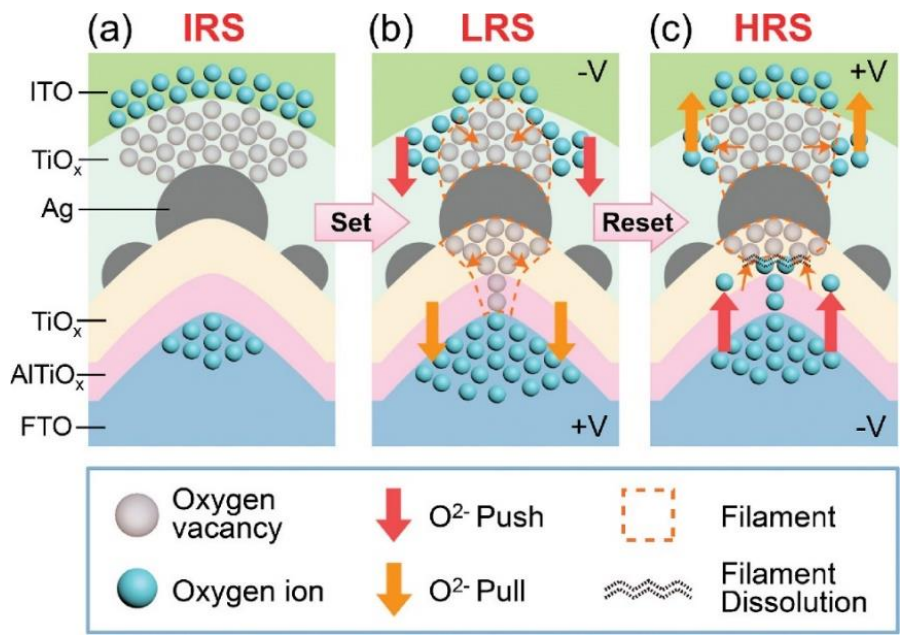

*Figure 9 Switching mechanism in a memory device showing SET/RESET performance. (need to take permission from F8)*

## 2.5 Compositions for RRAM device fabrications

Hickmott in 1960 showed the first experimental demonstration of the negative value of resistance in five anodic oxide compositions SiO, $Al_2O_3$, $Ta_2O_5$, $TiO_2$, and $ZrO_2$ in a MIM structure, which was further investigated by many researchers over the year. In 2008 the working of a memristor was first successfully demonstrated by the HP laboratories in 2008; since then, it has gained a significant amount of interest for future application purposes.

So far, the RS behavior has been explored in numerous classes of inorganic as well as organic material starting right from oxides, halides, chalcogenides, heterostructure, 2D material, 1D material, etc.[66]. In this review, we will be putting an insight into inorganic material potentially used for RRAMs and its derivative application purposes [67].

**Resistive switching in metal oxides:**

Among all available materials, the metal oxide enjoys a simple structure with many advantageous features such as wide bandgap, low preparation cost, well precise electrical behavior, electrically and thermally stable, electrochemically stable broad morphology, and safe for the environment. These fascinating properties make them an ideal candidate to be used for diverse application purposes such as sensors, optical waveguides, spintronics, transparent conducting oxides (TCO), photovoltaics, memory devices, and resistive switching. The first investigation of resistive switching behavior in binary oxide was studied by Hickmott in 1962 on $Al/Al_2O_3/Al$ device configuration by [68–70]. Since then, this phenomenon has been largely explored in numerous binary metal oxides such as ZnO, NiO, CuO, MgO, $HfO_2$, $TiO_2$, $Al_2O_3$, $Nb_2O_5$, $Fe_2O_3$, $Ta_2O_5$, and many more. Among all ZnO gets an edge for the same due to its fabrication-friendly nature [71]. Over the year, the researcher has studied ZnO widely for RRAM application purposes by employing expensive electrodes such as Ag, Au, and Pt, but Lin et al. has investigated the RRAM properties using inexpensive Aluminum electrode at different Argon/Oxygen concentration ($Ar/O_2$ ratio of 2 and 3). $Ar/O_2$ gas flow ratio of 2 displays an excellent switching cycle up to 221 times and a high HRS/LRS ratio, which reaches $10^9$ shown in **Fig.10** [72].

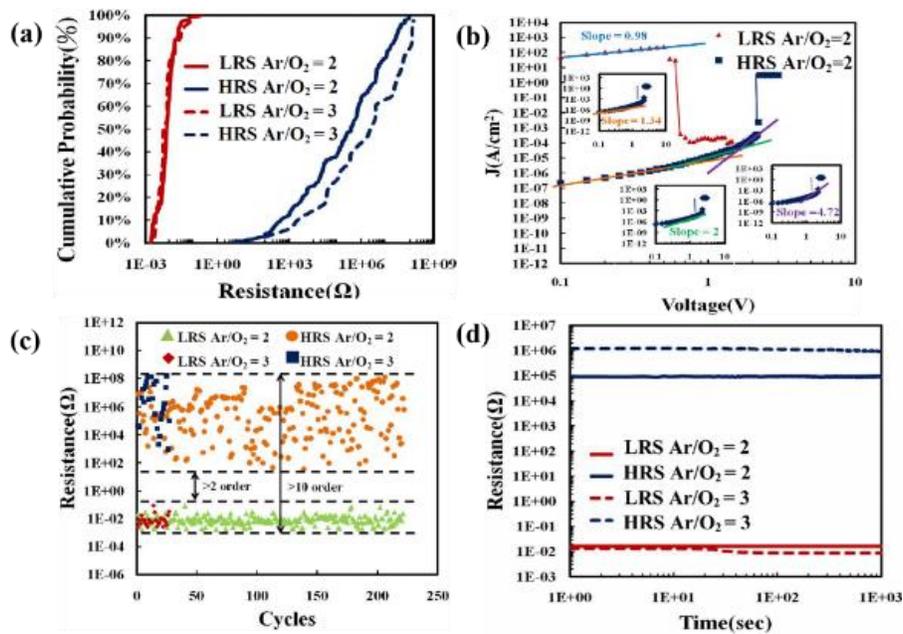

*Figure 10 I-V cure and switching cycle of (need to take permission from S12)*

The effect of doping on ZnO has been studied by numerous authors, [73–81] including, Boppidi et al. who has investigated the RS in Cu doped ZnO synthesized by low coast chemical process. It was revealed that the fabricated device displays resistive switching behavior, which usually comes into the picture due to the movement of mobile oxygen vacancy in the system. Further, there was enough room to enhance the property of the fabricated device, achieved by annealing the same at 800°C. An improved ferroelectric property along with the higher ON/OFF ratio, outstanding switching property, low coercive voltage, and low SET/RESET voltage of around 1.40 V was achieved in the annealed Cu: ZnO. This study enlightens the development of forming free low power NV-RAM [79]. Another doping-related study was performed by Xu et al. the group has investigated the RS behavior in Cr doped ZnO thin film deposited on Pt/Ti/SiO$_2$/Si/Pt substrates, with the increase in doping concentration (1% to 5%), the ratio of HRS to LRS also enhanced drastically from 17 to $7\times10^3$. The prepared device displays a bipolar resistive phenomenon without forming process, confirmed by AFM [76]. Saini et al. has explored the multilevel resistive switching behavior in Cu doped ZnO thin film, by incorporating the pulsed laser deposition (PLD) technique. The resistive switching measurement was performed in light and as well as in the dark and displayed ohmic conduction at a lower voltage (V< 1.15V), followed by tratrap-filledmited current in a voltage range of 1.15 V < V < 2.3 V and finally trap free space charge limited conduction (SCLC) current at voltage beyond 2.3 V for HRS in forward bias. Whereas, for LRS in forward bias, it displays an ohmic behavior till 0.75 V and shows a trap-free conduction beyond 1.65V. The PLD grown heterojunction device displays a forming free, multilevel RS memory operating on low voltages, which would be used as visible light light-controlled memristive devices, useful for low power operation [82]. ZnO can display various morphology such as nanowire, nanoforest, nanoribbon, Nanohelix, a nanobelt, and nanosheets whichaveas been used for RS application like Chen et al. has successfully deposited 1D nanowire structure on Ta$_2$O$_5$ substrate, the fabricated Au/p-n ZnO NWs/ Ta$_2$O$_5$ /Au display an excellent resistive property with a SET/RESET ratio of the order of $10^5$ at the reading voltage of 0.1 V. The Au/p-n ZnO NWs/ Ta$_2$O$_5$ /Au device configuration also shows the rectifying behavior with the reverse current ratio of $4.5\times 10^4$ at LRS. The fabricated device can be operated beyond 600 cycles, displaying its excellent endurance behavior, which can be a decisive factor in fabricating 5G memory device with artificial intelligence (AI) [83]. TiO$_2$ has also been extensively investigated by the researcher for memory application due to its easy fabrication process, high dielectric constant, stable crystal structure, and high reversible capacity [84–87]. Chen et al. has utilized TiO$_2$ as an active layer interceding between two conducting electrodes (ITO/TiO$_2$/Pt) for RRAM application. The Electrical measurement of the prepared RRAM memory device displays a notable low operational parameter with the SET and RESET voltage of 0.6 V and -0.5V, respectively. The TiO$_2$ based RRAM device displays a bipolar RS behavior with an HR to LRS resistance ratio of greater than one order of magnitude and a data retention period of something over $10^4$ seconds. The I-V curve on the log-scale reveals the conduction mechanism of the RRAM, mostly dominated by the ohmic conduction in SET as well as in RESET states [84]. Ismile et al. displayed a meliorate resistive switching behavior in a bilayer $TiO_2$ based memory device by inserting a switching layer of non-stoichiometric $CeO_{2-X}$ between the top TaN electrode and the $TiO_2$ layer in TaN/$CeO_{2-x}$/$TiO_2$/Pt RRAM device, deposited through radio frequency (RF) sputtering technique at 25 ºC. In contrast to single-layer switching devices, the fabricated bilayer device displays a significant improvement in DC over the switching cycle ($>1.2\times10^4$), retention period ($>10^4$s), switching voltage, reliability, and endurance. The enhanced switching uniformity in the $TiO_2$ based bilayer RS device is also achieved, the primary reason responsible for the same is oxygen vacancy assisted formation and destruction of conducting filaments [88]. A similar kind of enhanced resistive switching behavior was performed by Trapatseli et al. by depositing a very thin $Al_2O_{3-y}$ buffer layer between $TiO_2$ film and top contact. The inclusion of a buffer layer has two roles, first to increase the oxygen vacancy in the system and the other one was to provide a stable formation and eruption of conductive filament, which eventually increases the endurance and other device performance (switching speed, adaptability of switching, and low operational power, etc.). The application of this type of bilayer device is not just limited to NVRAMs application but has also proved to be an excellent candidate to be used for neuromorphic computing (detailed discussion is provided in the next section) [89]. Gu et al. has probed the effect of oxygen vacancy on the resistive switching behavior in epitaxially grown $TiO_2$ thin film on conducting Nb-$SrTiO_3$ substrates through RF-magnetron sputtering. One can easily observe a transition of resistive switching mode from Valance charge metallization (VCM) to electrochemical metallization (ECM) type, with a noticeable decrease in the concentration of oxygen vacancies in the system. ECM type resistive switching behavior is infrequent to observe in $TiO_2$, which further improves the memory device's performance by enhancing the switching ratio ($>10^5$), reducing the leakage current in HRS, and inducing specific quantized conductance in LRS. This work provides a concert proof of modulating the properties of ReRAMs by playing with defect engineering [90]. A unique kind of memristor device was fabricated by Jaffar et al. by using nanocomposite resistive switching material consisting of photoactive, azopolymer matrix (PDR1A), and $TiO_2$ nanorods (PDR1A/$TiO_2$ NRs). The fabricated device displays outstanding memristor switching properties along with the reversible polarization-driven optical switching behavior arising from the photoactive nature of PDR1A. The PDR1A/TiO2-NRs based fabricated memristor usually displays a higher nonlinear switching behavior which was further probed by employing the charge-flux memristor model [91]. Michalas et al. has studied the role of various top contact on the nature of resistive switching on a prototypic device made using gold as a bottom electrode, $TiO_2$ has been used as an active layer, and three different top contacts Nickel (Ni), Gold (Au), and Platinum (Pt) have been used in this work. The temperature-dependent I-V measurement in a non-switching regime reveals a bias-induced interfacial resistive switching mechanism, which is well supported by the steady-state SET and RESET process. A typical bipolar switching behavior was displayed by the device having top contact with Ni, Pt, and Au [91]. Ebenhoch et al. has explored the memristive behavior in hydrothermally synthesized $TiO_2$ nanorods array placed between FTO acting as a bottom electrode and Au as a top electrode (FTO/TiO$_2$/Au). A significant decrease in resistance of the bulk nanorods was observed, mainly arising due to the accumulation of mobile oxygen vacancies at the interface of FTO and grain boundaries. A relaxation behavior in the fabricated device was probed by the I-V measurement. However, the fabricated TiO$_2$ NRs based memristor device displays a hysteresis behavior but does not supervene the typical unipolar, bipolar or complementary switching path. Instead, it displays a time-dependent current-voltage relation. Further, the effect of defect reaching grain boundary on the I-V behavior was studied by coupling the electronic measurement with SEM and STM [91]. TiO$_2$ NRs can be prepared by numerous methods, among all the low-cost hydrothermal method was incorporated by Huang et al. to synthesize TiO$_2$ NRs and explored the RS behavior in it. Electrical measurement was performed on FTO/TiO$_2$ NRs/Pt RRAM device, revealing the nonlinearity up to 10, which eventually decreases the leakage current up to $10^{-4}$ Acm$^{-2}$, thus endorsing the self-selecting RS behavior. the resistance of HRS and LRS are predicted using I-V measurement, and it is of the order of $10^5$ and $10^4$ at the read voltage of 6V, and the Ratio of $R_{HRS}/R_{LRS}$ of approximately 10 to $10^2$ respectively. The obtained results offer a great opportunity to utilize the self-selecting RS behavior in a binary metal oxide, which is much needed to overcome the sneak path issue in the RRAM element to attain large passive crossbar arrays [92].

Recently, HfO$_2$ has also been vigorously studied for the RS application due to its low operation power, high nonlinearity, and high speed [93]. Luo et al. has proposed a unique solution to solve the current issue in a 3D vertical crossbar array by fabricating a forming free self-rectifying resistive switching memory device by utilizing the HfO$_2$ in Pd/HfO$_2$/WOx/W heterostructure. The fabricated memory device shows a high uniformity, high operational speed (100 ns), a low switching voltage of less than 3V, low read voltage of $10^9$ order, and sturdy reliability, thus presenting the HfO$_2$ heterostructure device as a quiescent candidate to be used for high-density RS applications [94]. Wang et al. has demonstrated the RRAM property in the trilayer structure of Al$_2$O$_3$/HfO$_2$/Al$_2$O$_3$, deposited on TiN-coated Si substrate through atomic layer deposition (ALD) technique. The trilayer Pt/Al$_2$O$_3$/HfO$_2$/Al$_2$O$_3$/TiN/Si device displays an enhanced resistive switching behavior, possibly due to the presence of two interfacial layers between Al$_2$O$_3$/I.L./HfO$_2$/I.L./Al$_2$O$_3$. The fabricated memory device demonstrates a bipolar resistive feature with the resistance ratio under 10 for RESET/SET state and the data retention period of more than 10 years kept at 85$^o$C, thus can be used as an oxide-based memory device. Ryu et al., for the first time, demonstrated the erase-free, high reliable, and energy-efficient multibit RS phenomena in W/HfO$_2$/TiN devices [95]. Zhang et al. has fabricated a flexible memory device by utilizing a bilayer of TiO$_2$/HfO$_2$ on a polyethylene naphthalate substrate, displaying phenomenal scalability, high durability, and outstanding mechanical flexibility. A very meager fluctuation has been observed upon I-V measurement in LRS and HRS, indicating outstanding endurance behavior and almost no deterioration in electrical performance has been observed under the application of mechanical stress with the bending radius changing from 10 to 70 mm, thus can create a boom in the field of wearable electronics [96]. While resistive switching behavior has been observed in a ample range of transition metal oxides (TMOs), among them, the Niobium pentoxide ($Nb_2O_5$) displays metal-insulator-metal (MIM) type two-terminal memristor application in both crystalline and as well as in amorphous phases exhibiting unipolar, bipolar and non-polar RS mode. The primary reason behind this is its simple structure, wide bandgap (> 3.5eV), high permittivity (40), higher capacitance density (17 fF/µm$^2$), and assorted switching characteristics. However, its low SET/RESET ratio and forming process somehow limits its practical applications [97–105]. The RS behavior in Pt/$Nb_2O_5$/Al device was shown by Deswal et al., synthesized through the reactive sputtering method. The fabricated device displays a nonvolatile unipolar resistive switching behavior with the SET/RESET resistance ratio greater than 10$^3$. I-V measurement reveals the Ohmic and SCLC type conduction mechanism in LRS and HRS, respectively. The author incorporated the oxygen ion vacancy model to study the formation and rupturing of the conductive filament during the ON and OFF state, which will eventually help in the fabrication of a highly competent RS device [106]. Deposition temperature also plays an important role in the RS mechanism. Its effect on the PLD deposited $Nb_2O_5$ films on Silicon substrate was studied by Xu et al. with the increase in the in situ deposition temperature from 250 ºC to 400 ºC, a drastic increase in the concentration of oxygen vacancy in Cu/ $Nb_2O_5$/Pt memory devices was observed which was also confirmed by the XPS measurement. With the rise in the concentration of oxygen vacancy, the resistive switching in the SET process undergoes a transition from abrupt to stepwise switching. In contrast, the RESET process changed from abrupt to gradual switching for the Cu/$Nb_2O_5$/Pt devices, thus providing a novel route to tailor the resistive switching behavior of TMO-based RRAMs by varying the concentration of oxygen vacancy during the deposition process [99]. Sahoo et al. have observed the room temperature bipolar resistive switching behavior in Pt/$Nb_2O_5$/Pt and Bare conductive paint BCP $Nb_2O_5$/Pt device fabricated through e-beam deposition technique. The fabricated device displays a replicable switching behavior at a very low reading voltage of 0.1 V with a compliance current of around 6 mA and also depicts an excellent data retention ability. The reliability test of $Nb_2O_5$ based memristive device stack structures has been examined by inspecting the retention ability (up to 600 min) and endurance test of over 1000 cycles, which proved to be way better than the ssingle-layerNb$_2$O$_5$ based memristor device [107]. Zhou et al. has shown the unique kind of electrode effectuate polarity switching in $Nb_2O_5$/$NbO_x$ resistive switching devices deposited through reactive magnetron sputtering. Here the $Nb_2O_5$ and $NbO_x$ play the role of switching layer and oxygen vacancy reservoir (Vo), embracing a CF formed during electroforming. Two types of devices have been fabricated with different top electrode compositions, Pt/ $Nb_2O_5$/ $NbO_x$/Ru and W/ $Nb_2O_5$/ $NbO_x$/Ru, respectively. The device with Pt electrode displays a clockwise RS switching while the device with W electrode depicts a counter-clockwise RS switching. Thorough investigations suggest that reactive electrodes deliver an oxygen ion reservoir near the active region of the CFs and it is well known that the oxygen reservoir controls the formation and rupture of the CFs through interfacial redox reaction, thus kicks the transformation of switching polarity. Despite having antagonistic switching, both the devices show ohmic conduction for LRS and Schottky emission for HRS. Detailed investigation for the same has been performed using Schottky emission fitting and provides an exciting result. This work shows a unique way of tailoring and enhancing the RS device's performance using an oxygen ion reservoir [105]. Nath et al. have shown enhanced device reliability and resistive switching in undoped $Nb_2O_5$ and Titanium doped $Nb_2O_5$ thin film-based cross-point devices fabricated through RF sputtering method. A significant variation in the device characterizes observed for the $TiO_2$ and $Ti/Nb_2O_5$ devices when the Ti/Nb fraction is 0. 71, the device possesses a low resistance and displays a continuous I-V relation, on the other hand, when the Ti/Nb fraction $\leq 0.31$, demonstrate an NDR behavior with monotonous I-V characterizes. The impact of Ti doping on the reliability and uniformity of the device was improved by the Ti doping and can withstand more than 6000 voltage-controlled switching cycles. The fabricated device was further tested for temperature-dependent I-V measurement, and lumped-element modeling reveals that the NDR comes into the picture due to changes in the oxide conductivity. This study further delivers a new prospect for designing and developing RS devices for particular applications [107]. As discussed above, doping can enhance the material's physical, chemical and electrical property, which the Sedghi et al. has also observed by doping the $Ta_2O_5$ by fluorine. The doping has drastically improved the stability of the resistive state, and as well as endurance, which drastically escalated from 50,000 cycles for undoped $Ta_2O_5$ to $10^6$ for F doped $Ta_2O_5$ memory devices, possible reason behind this can be the easing out of the oxygen vacancies from the $Ta_2O_5$ system. Density functional theory (DFT) reveals that the doping of fluorine in a two-fold O vacancy site in $Ta_2O_5$ reduces the total quantity of the defects, which were capable enough to form an alternative conducting path, thus enhancing the current stability in the SET states. More than two years of retention period has been confirmed from the Stress voltage accelerated retention failure method [108]. The same group has checked the effect of nitrogen doping on the data retention period, stability and memory window of resistive state switching of $Ta_2O_5$ based RS device fabricated through atomic layer deposition (ALD). The incorporation of nitrogen reduces the electronic defect level, which eventually permits the CFs to form with much stable resistance which is well suited for MLC switching [109]. Chang et al. has shown a unique kind of work by showing the effect of humidity on single and bilayer $Ag/Ta_2O_5/Fe_3O_4/Pt$ and $Ag/Fe_3O_4/Ta_2O_5/Pt$ structures. The switching behavior of the fabricated single and bilayer devices was investigated in the vacuum and as well as in atmosphere conditions, which reveals that the development of silver filament was challenging in low humidity regions, which further led to an increase in SET voltage. Moreover, the bilayer RRAM displays an enhanced switching voltage, endurance, reliability, low forming voltage, and high endurance compared to the single-layer devices. This study provides an edge to bilayer structure RRAMs due to their unpretentious structure and low fabricating cost to expand the design and development of future memory devices [110]. Lee et al. has inspected the role of a top metal electrode (Ti, Ag) on the switching characteristics and synaptic behavior. Device with Ti as the top electrode needs an electroforming process to start the switching process, whereas the device with Ag as the top electrode doesn't require a forming process to do the same in RESET process, primary reason behind this is the difference in the composition of the conductive filament of both devices. Similarly, the device with Ti as a top electrode displays a VCM type switching mechanism, on the other hand, the Ag as a top electrode displays both VCM and ECM type dual-mode switching mechanism. This type of study enhances the level of understanding of the complex switching mechanism of the RRAM device, which will further fill the technical void currently faced by the researcher working in the field of memory devices [111]. Song et al. has shown analog resistive switching behavior in the bilayer of TiW/$Al_2O_3$/ $Ta_2O_5$5/Ta stack deposited through ALD method. The current conduction mechanism of the fabricated devices was dominated equally by both Poole Frenkel (P-F) emission and as well as Schottky emission. An abundant amount of oxygen vacancy was found on the device's interface, which plays a critical role in the modulation of the Schottky and tunneling barriers. A well-defined bipolar RS with multiple states and analog switching behavior was demonstrated by the interfacial device. Multiple states were confirmed by applying 128 successive indistinguishable pulses of 20 us in depression and 600 ns in potentiation and revel the data retention period of $10^4$ s for the multiple states [112].

**Perovskite material for memory device application**

For the last couple of decades, the perovskite material known to be the most probable candidate for industrial application owing to its unique features such as widbandgapap, high curie temperature, chemically stable, stable crystal structure, and high tolerance ratio, and susceptibility to doping, etc. Additionally, the strongly coupled electrons in these systems offer various electronic phases and display multifunctionality in behavior such as electroresistance, colossal magnetoresistance, superconductivity, ferroelectricity, and ferroelasticity, multiferroicity, etc. [113–123]. Also, the metal-insulator transition displays a substantial amount of variation in resistance to applying even a minute electrical stimulus, making them an eligible candidate to be used for RRAM application [122]. Among all, $BaTiO_3$ is the most common perovskite material explored for resistance switching applications because of its eco-friendly nature, high dielectric constant, remarkable polarization, low dielectric loss, and the much-needed transition temperature above room temperature. Enhanced resistive switching was demonstrated by Silva et al. by fabricating the set of RS devices by utilizing ferroelectric $BaTiO_3$ (BTO) material and inserting $HfO_2$:$Al_2O_3$ (HAO) layer between the top electrode Pt and insulating layer BTO as in Pt/BTO/ITO and Pt/ HAO/BTO/ITO. Both devices show a decent ferroelectric behavior with a remnant polarization of 8.9μC/$cm^2$ and 5.4 9μC/$cm^2$. The relation between resistive switching behavior and ferroelectric polarization reversal has been explored at various temperatures and reveals an enhanced RS switching with the RS ratio of $10^6$, which normally arises due to the formation of barrier variation in Pt/ HAO/BTO/ITO device, which was suppressed while approaching the curie temperature (Tc). An improved endurance behavior of $10^3$ was shown by the Pt/ HAO/BTO/ITO device after $10^9$ switching cycle, which makes them eligible candidates to develop a ferroelectric driven RS memory device [124]. Another investigation in this regime was performed by Li et al. exploring the nonvolatile RS behavior by controlling the direction of polarization of BaTiO3 (BTO) layer in BaTiO$_3$/La$_{0.7}$Sr$_{0.3}$MnO$_3$ (BTO/LSMO) layered heterostructure grown on SrTiO$_3$ (STO) substrate through PLD. The dielectric property of the BTO layer has been measured and reveals a much lesser value, possible reason behind this is the accumulation of oxygen vacancy at the interface of LSMO and BTO, which usually arises due to lattice mismatch at the time of deposition. The temperature-dependent resistivity metal-insulator transition temperature of the BTO/LSMO heterostructure has been investigated in this work, and it reveals that the transport property of the LSMO film is sensitive to the external electric field. When a positive voltage bias was applied to the BTO layer, it enhances the resistivity of the heterostructure due to the formation of an oxygen-deficient layer (transfer of oxygen vacancy from BTO/LSMO boundary to the LSMO field); however, a reduction in resistivity was detected on applying a negative bias. Obtained results display the resistive switching behaviour in the multiferroic heterostructure, which can be further utilized to develop magnetic storage devices [125]. An important study on perovskite oxide-based RRAM application was performed by Zheng et al. by fabricating a multilayer RRAM device in PbZr$_{0.52}$Ti$_{0.48}$O$_3$/La$_{0.67}$Sr$_{0.33}$MnO$_3$/Nb: SrTiO$_3$ (PZT/LSMO/NSTO) heterostructures, where the thickness of ferroelectric LSMO was varied. Electrical measurements show that the ferroelectric property of the PLD grown heterostructure device doesn't fluctuate much by changing the thickness of the LSMO layer. At the same time, the resistive switching behavior is meticulously related to the LSMO thickness. When the thickness of LSMO was kept at 18 nm, a switching ratio of about $10^3$ was achieved in the fabricated heterostructure device. A deterioration in switching property was observed when the thickness of the ferroelectric layer was 7 and 70 nm. Detailed investigation of energy band structure and I-V curve states that the RS behavior strongly depends upon hustling between the tunability of the width of a depletion layer and the capability of ferroelectric filed effect, thus providing an efficient way to tune the RS behavior of multilayer heterostructure device by inserting a ferroelectric buffer layer [125]. lee et al. have explored the RS phenomena in ALD prepared SrTiO$_3$ perovskite material sandwiched between Pt-based electrodes. The fabricated device was susceptible to the electroforming process only when a negative bias voltage was applied, whereas it displays a bipolar RS behavior with set and reset in positive as well as in negative biases, respectively [125]. Khalid et al. explored the RRAM application in hybrid nanocomposite devices comprised of SrTiO$_3$ and PVA, sandwiched between conducting Silver electrodes. The fabricated device displays a bipolar RS behavior with low operating voltage, outstanding data retention period of $8.6 \times 10^4$, switching ratio of 30 at the read voltage of 1V, and electrical endurance of 500 operating cycles. Along with the switching behavior, the mechanical robustness of the device was also tried and tested and revealed the mechanical endurance of 500 bend cycles at the bending diameter of 15 mm, which makes them a potential candidate to be used as a flexible device [125]. Apart from BaTiO$_3$ another perovskite material that has been explored for a similar application purpose is BiFeO$_3$ as it displays an excellent magnetodielectric property;

Kumari et al. has fabricated a RRAM device using sol-gel prepared $BiFeO_3$ thin film sandwiched between Ag and FTO conducting electrode ($Ag/BiFeO_3/FTO$). The fabricated device displays a stable bipolar RS behavior with the ON/OFF ratio of 450, low operating voltage of 1.1 V for SET and -1.5 for RESET state, respectively and large data retention period upto $10^4$ sec. The possible conduction mechanism in LRS and HRS is governed by the ohmic and trap assisted space charge limited current (SCLC) in both positive and negative bias [126]. Same group has shown the forming free switching in $BiFeO_3$ based RRAM device grown on glass coated FTO as a bottom electrode and Aluminum as a top electrode. The fabricated device displays a nonvolatile bipolar switching behaviour with DC and AC endurance of more than 250 and 7100 cycles, respectively. The data retention period is similar to the previous one ($10^4$ sec.). The switching mechanism was governed by the rupture and formation of oxygen vacancy intermediated conducting filaments, endorsed by the AlOx layer formation [127]. Yan et al. has shown the capacitance and resistance-based memory device by utilizing the multilayer stacking of $DyMnO_3$ and Nb: $SrTiO_3$ perovskite oxide stacked between conducting Au electrode and staking will look like $Au/DyMnO_3/$ Nb: $SrTiO_3/Au$. The device shows a very high retention time of $10^5$ sec with a ratio of $R_{LHS}/R_{HRS}$ is greater than 100 over the $10^8$ switching cycles. The serial connected $Au/DyMnO_3/$Nb: $SrTiO_3$ heterostructure acts as a high nonlinear resistor paralleling with a capacitor and improves the stability of the memory device [127].

## Conclusion and outlook

From the discovery of the first primordial computer in the 1970s to the modern day's supercomputers, smartphones, laptops, and automated vehicles, the most critical and common part that control its operation is its nonvolatile memory. This review deals with a concise introduction to the growth and development of memory architecture, the operating mechanism, and the limitations while keeping a valuable insight into the area of evolving memory technologies. The nonvolatile memory has been differentiated into PCM, STTRAM, SRAM, DRAM, FLASH memory, and RRAM, among all the RRAM, proves to be the most promising one due to its unique properties such as high fatigue resistance high polarization switching ($10^{12}$ cycles), low operating voltage (within 1V), ultrafast processing speed (less than one ns), and can be further scaled down to compact size, thus can be effectively used as the next generation Nonvolatile memory technologies. This review extensively deals with the eligible oxide-based compositions such as ZnO, $TiO_2$, $HfO_2$, $Ta_2O_5$, $Nb_2O_5$, and many more for the RRAM applications and the effect of doping, defect engineering, varying the oxygen content, and impact of different electrode material on the performance of the RRAM device has also been discussed in detail. The low power and efficient way to integrate a large number of the incoming signal of RRAM makes them eligible for bio-inspired based artificial synaptic applications such as neuromorphic computing, artificial intelligence, internet of things, etc. discussed in this review. Although a substantial amount of success has been attained in RRAM technology; but still much work is needed as RRAM still mimicking from numerous challenges in form of higher operation current, reduced resistance ratios, variability problems, reliability at cell-level, and device-level, and, highly smooth structure designs which must be encountered for its better and smooth application in diverse field.

**Acknowledgments**: RG acknowledges IIT Kanpur for a senior research fellowship. PCM acknowledges the Department of Science and Technology for a start-up research grant (SRG/2019/000391), IIT Kanpur for initiation and special grant (IITK/CHM/2019044), and the Council of Scientific & Industrial Research (CSIR, Sanctioned NO.:01(3049)/21/EMR-II), New Delhi.

[31] Y. Li, S. Long, Y. Liu, C. Hu, J. Teng, Q. Liu, H. Lv, J. Suñé, M. Liu, Conductance Quantization in Resistive Random Access Memory, Nanoscale Research Letters. (2015). https://doi.org/10.1186/s11671-015-1118-6.

[32] M. Lee, S. Han, S.H. Jeon, B.H. Park, B.S. Kang, S. Ahn, K.H. Kim, C.B. Lee, C.J. Kim, I. Yoo, D.H. Seo, X. Li, J. Park, J. Lee, Y. Park, Electrical Manipulation of Nanofilaments in Transition-Metal Oxides for Resistance-Based Memory, (2009).

[33] Y. Huang, T. Shen, L.-H. Lee, C. Wen, S. Lee, Low-power resistive random access memory by confining the formation of conducting filaments, AIP Advances. 6 (2016) 065022. https://doi.org/10.1063/1.4954974.

[34] G. Baek, M.S. Lee, S. Seo, M.J. Lee, D.H. Seo, D. Suh, J.C. Park, S. Park, H.S. Kim, I.K. Yoo, U. Chung, I.T. Moon, Highly Scalable Non-volatile Resistive Memory using Simple Binary Oxide Driven by Asymmetric Unipolar Voltage Pulses, (n.d.) 587–590.

[35] R. Waser, M. Aono, Nanoionics-based resistive switching memories, (2007) 833–840.

[36] B.J.C. Scott, L.D. Bozano, Nonvolatile Memory Elements Based on Organic Materials **, (2007) 1452–1463. https://doi.org/10.1002/adma.200602564.

[37] A. You, M.A.Y. Be, I. In, Investigating the switching dynamics and multilevel capability of bipolar metal oxide resistive switching memory, 103514 (2015) 98–101. https://doi.org/10.1063/1.3564883.

[38] M.M. Rehman, H.M.M.U. Rehman, J.Z. Gul, W.Y. Kim, K.S. Karimov, N. Ahmed, Decade of 2D-materials-based RRAM devices: a review, Science and Technology of Advanced Materials. 21 (2020) 147–186. https://doi.org/10.1080/14686996.2020.1730236.

[39] L. Liu, C. Liu, L. Jiang, J. Li, Y. Ding, S. Wang, Y.G. Jiang, Y. Bin Sun, J. Wang, S. Chen, D.W. Zhang, P. Zhou, Ultrafast non-volatile flash memory based on van der Waals heterostructures, Nature Nanotechnology. 16 (2021) 874–881. https://doi.org/10.1038/s41565-021-00921-4.

[40] J.S. Lee, Progress in non-volatile memory devices based on nanostructured materials and nanofabrication, Journal of Materials Chemistry. 21 (2011) 14097–14112. https://doi.org/10.1039/c1jm11050k.

[41] A. Architectures, Resistive Random Access Memory ( RRAM ), n.d.

[42] A. Chen, Solid-State Electronics A review of emerging non-volatile memory ( NVM ) technologies and applications, Solid State Electronics. 125 (2016) 25–38. https://doi.org/10.1016/j.sse.2016.07.006.

[43] D.C. Gilmer, G. Bersuker, Fundamentals of Metal-Oxide Resistive Random Access Memory ( RRAM ), (n.d.) 71–92.

[44] S. Resistance, Y. Syu, G. Chang, T. Chu, G. Liu, Y. Su, M. Chen, J. Pan, Characteristics and Mechanisms of Random Access Memory, 34 (2013) 399–401.

[45] M. Lanza, H.P. Wong, E. Pop, D. Ielmini, D. Strukov, B.C. Regan, L. Larcher, M.A. Villena, J.J. Yang, L. Goux, A. Belmonte, Y. Yang, F.M. Puglisi, J. Kang, B. Magyari-köpe, E. Yalon, A. Kenyon, M. Buckwell, A. Mehonic, A. Shluger, H. Li, T. Hou, B. Hudec, D. Akinwande, R. Ge, S. Ambrogio, J.B. Roldan, E. Miranda, J. Suñe, K.L. Pey, X. Wu, N. Raghavan, E. Wu, W.D. Lu, G. Navarro, W. Zhang, H. Wu, R. Li, A. Holleitner, U. Wurstbauer, M.C. Lemme, M. Liu, S. Long, Q. Liu, H. Lv, A. Padovani, Recommended Methods to Study Resistive Switching Devices, 1800143 (2019) 1–28. https://doi.org/10.1002/aelm.201800143.